\documentclass[prd,twocolumn,superscriptaddress,nofootinbib,letterpaper,altaffilletter]{revtex4}
\usepackage{latexsym}
\usepackage{hyperref}
\usepackage{graphicx}
\usepackage{color}
\usepackage{times}
\usepackage{fancyhdr}
\usepackage{float}
\usepackage{ulem}
\makeatletter
\makeatletter
\def\@fnsymbol#1{\ifcase#1\or * \or  $+$ \or  \$ \or \#  \or \dag \or \ddag \or
$\mathsection$ \or $ \mathparagraph$ \or $\|$  \or \textordfeminine \or \textbul
let   
\or ** \or $++$ \or  \$\$ \or \#\#  \or \dag\dag \or \ddag\ddag \or
$\mathsection\mathsection$ \or $ \mathparagraph\mathparagraph$ \or $\|\|$  \or 
\textordfeminine\textordfeminine \or \textbullet \textbullet \or *** \or $+++$ 
\or  \$\$\$ \or \#\#  \or \dag\dag \or \ddag\ddag \or
$\mathsection \mathsection\mathsection$ \or $ \mathparagraph 
\mathparagraph\mathparagraph$ \or $\|\|\|$  \or 
\textordfeminine\textordfeminine\textordfeminine \or 
\textbullet\textbullet\textbullet \or \else \@ctrerr\fi}
\makeatother

\newcommand\rhom{\ensuremath{\rho_{m}}}
\newcommand{\deff}{\ensuremath{D_{\mathrm{eff}}}}
\newcommand{\dchirp}{\ensuremath{D_{\mathrm{c}}}}
\newcommand\half{\mbox{$\frac{1}{2}$}}
\newcommand{\cl}{\ensuremath{\mathcal{C}_{L}}}
\newcommand{\mchirp}{\ensuremath{\mathcal{M}}}
\newcommand{\lamvec}{\ensuremath{\vec{\lambda}}}

\def\thercsid{\relax}
\def\rcsid#1{\def\next##1#1{\def\thercsid{##1}}\next}
\rcsid$Id: systematics.tex,v 1.41 2007/07/13 02:16:25 patrick Exp $

\renewcommand{\today}{\number\day\space\ifcase\month\or
  January\or February\or March\or April\or May\or June\or
  July\or August\or September\or October\or November\or December\fi
  \space\number\year}

\begin{document}

\title{Interpreting the results of searches for gravitational waves from
coalescing binaries}

\author{Stephen Fairhurst}
\affiliation{LIGO - California Institute of Technology,
Pasadena, CA  91125, USA\\
School of Physics and Astronomy, Cardiff University,
Cardiff, CF2 3YB, United Kingdom}
\author{Patrick R. Brady}
\affiliation{University of Wisconsin-Milwaukee, Milwaukee, WI  53201, USA}

\date { RCS \thercsid; compiled \today }

\begin{abstract}

We introduce a method based on the loudest event statistic to calculate
an upper limit or interval on the astrophysical rate of binary
coalescence.  The calculation depends upon the sensitivity and noise
background of the detectors, and a model for the astrophysical
distribution of coalescing binaries.  There are significant
uncertainties in the calculation of the rate due to both astrophysical
and instrumental uncertainties as well as errors introduced by using the
post--Newtonian waveform to approximate the full signal.  We catalog
these uncertainties in detail and describe a method for marginalizing
over them.  Throughout, we provide an example based on the initial LIGO
detectors.

\end{abstract}

\maketitle
\section{Introduction}
\label{sec:intro}

The first generation of gravitational wave interferometric detectors
have achieved, or are approaching, their design sensitivities
\cite{Waldman:2006, Acernese:2006bj, Hild:2006bk}.  One of the most
promising sources of gravitational waves for these detectors are those
emitted during the coalescence of a binary system composed of neutron
stars or black holes.  The initial detectors will be sensitive to binary
neutron star (BNS) coalescences as far away as the Virgo cluster, while
for binary black holes (BBH) the reach is as great as 100 Mpc.  Thus,
with a year of data, there is a possibility of detecting gravitational
waves from these sources.  Even in the absence of a detection, the upper
limits which will be placed on the rates of binary coalescences will
begin to be astrophysically interesting.  Following the initial
searches, the detectors will be upgraded to enhanced and advanced
configurations where the sensitivity will increase by factors of around
two and ten respectively over the initial detectors.  In this paper we
describe a method which can be used for providing an astrophysical
interpretation of the results of a search for compact binary
coalescence.  In the absence of a detection, this will result in an
upper limit on the rate of such events.  If a signal is observed, then a
rate interval will be calculated.  The rate limit calculation makes use
of the loudest event statistic, first described in \cite{loudestGWDAW03}
and elaborated upon in \cite{ul}.  

The rate of gravitational waves observable at the detectors depends upon
the detector sensitivity and also the astrophysical model for the source
in question.  The standard assumption \cite{Phinney:1991ei} is that the
rate of binary coalescence in a galaxy follows the blue light
luminosity, which depends upon the star formation rate.  At cosmological
distances, it is appropriate to treat the universe as homogeneous and
isotropic.  However, for the initial and enhanced gravitational-wave
detectors, local anisotropies can have a significant effect.  In
particular, for initial LIGO and Virgo, a significant fraction of the
available luminosity comes from the Virgo cluster \cite{Nutzman:2004}.
Therefore, it is critically important to generate an accurate
distribution of the blue light in the local universe.  In
Ref.~\cite{LIGOS3S4Galaxies} a catalog of local galaxies has been
constructed for precisely this purpose.  In this paper, we describe in
detail how this galaxy catalog can be used to calculate the total
luminosity to which a gravitational-wave detector is sensitive.  In
addition to the galaxy distribution, this is dependent upon evaluating
the ability of a search to detect a binary with given parameters.   We
describe how this search efficiency can be calculated and folded
together with the galaxy distribution to obtain a total luminosity for a
given search. 

The amplitude of the gravitational-wave signal emitted by a coalescing
binary is dependent upon the component masses of the binary.  Thus, a
rate calculation (upper limit or interval) obtained from a search of
gravitational-wave data will be sensitive to the astrophysical model for
the distribution of the component masses of the binary.  Although
several neutron stars in binaries, and many isolated neutron stars, have
been observed as pulsars, there is still significant uncertainty in the
mass distribution \cite{Belczynski:2002}.  Furthermore, given the lack
of observed neutron star--black hole or binary black hole systems, the
distribution of component masses of these systems is highly uncertain.
With this in mind, we describe a simple method whereby mass dependent
rates can be obtained.  

In order to claim the detection of gravitational waves, it is vitally
important to have a good understanding of the distribution of events
which are due to instrumental noise.  The output from a search for
gravitational-wave transients is a list of candidate events which
survive all thresholds applied during the search.  These candidate
events can contain both background, noise events as well as true
gravitational-wave signals.  Without a good understanding of the noise
background, it will be impossible to determine that an event is due to a
gravitational-wave signal.  We shall see in detail later that a good
estimate of the noise background is equally important for calculating
rate limits or intervals.  

There are numerous uncertainties affecting our understanding of the
astrophysics, instrumental sensitivities and background estimates
required in interpreting the results.  The uncertainties in the galaxy
distribution are detailed in \cite{LIGOS3S4Galaxies}.  While the sky
position of nearby galaxies is well known, their distance from the earth
and blue light luminosity is measured less accurately.  The search for
gravitational waves from coalescing binaries makes use of waveform
templates calculated using the post--Newtonian approximation to general
relativity.  This will lead to some discrepancy between the waveform
used in the search and true physical waveform, particularly close to
coalescence.  In addition, the spin of the binaries is neglected in
current searches \cite{LIGOS3S4all}.  However, neutron stars and black
holes occurring in binary systems are expected to be spinning.  The
background of noise events is estimated by time shifting the data from
each instrument relative to the others.  This provides a reasonable, but
not perfect estimate of the background rate.  Furthermore, instrumental
calibration uncertainties will affect the reported sensitivity of the
instrument.  All of these will have an effect on the calculated rates.
We provide a detailed discussion of these systematic uncertainties and
describe a method by which these can be incorporated into the final rate
statements.  

The layout of the paper is as follows.  First, in Section
\ref{sec:inspiral} we briefly review the techniques used in searching
for gravitational waves from coalescing binaries.  In Section
\ref{sec:upper_limit} we discuss the loudest event statistic, and
demonstrate its application to a search for binary inspiral.  In Section
\ref{sec:systematics} we describe the various systematic effects which
can effect the rate upper limit, and describe a method for marginalizing
over these uncertainties.  Finally, in Section \ref{sec:summary} we
summarize the results.

Throughout the paper, the methods discussed are applied to an
illustrative example.  This example involves simulated results from two
detectors operating at the initial LIGO sensitivity for one year.  More
concretely we choose instruments with a binary neutron star horizon
distance (the distance at which an optimally oriented and located system
would produce a signal to noise ratio (SNR) of $8$ in a single detector)
of $35$ Mpc.  This is consistent with sensitivities achieved by the
Hanford and Livingston 4km detectors during the latter parts of the S5
science run.  Furthermore, we assume that the detectors and search are
such that the SNR threshold can be set at $5.5$ and noise background
will produce an expected rate of one event per year at a combined SNR of
$10$, or equivalently about $7$ in each instrument.  This value is
somewhat greater than the combined SNR of $8$ often assumed (see for
example \cite{Cutler:1992tc}) and reflects a realistic value given the
non-stationarity of the initial detectors.  Although the example
presented here involves simulated results, the methods are easily
generalized to searches of real data.  Indeed, the reported upper limits
from searching the data from the third and fourth LIGO science runs
\cite{LIGOS3S4all} were obtained using the methods described here.

\section{Details of inspiral searches}
\label{sec:inspiral}

The gravitational waveform emitted by a coalescing binary can be
calculated to high accuracy within the post--Newtonian expansion of
general relativity.  In Section \ref{sec:waveform}, we sketch the
details of the inspiral waveform, paying particular attention to those
quantities which have a direct impact on the detectability of the
waveform at a given detector.  Then, in Section \ref{sec:match_filter},
we describe the matched-filter search method which is used in searching
for the waveform.  We finish with a brief description of the extension
to multi-detector searches.

\subsection{The waveform}
\label{sec:waveform}

\noindent The gravitational-wave strain incident on an interferometer is
given by
\begin{equation}\label{eq:ht}
  h(t) = F_{+}(\theta,\phi,\psi) h_{+}(t) + 
  F_{\times}(\theta,\phi,\psi) h_{\times}(t) \, ,
\end{equation}
where $F_{+}$ and $F_{\times}$ are the detector response functions and
$h_{+}$, $h_{\times}$ are the two polarizations of the gravitational
wave.  The detector response functions depend upon the sky location
$(\theta, \phi)$ and polarization angle $\psi$ of the source relative to
the detector according to \cite{Anderson:2000yy}
\begin{eqnarray}\label{eq:det_response}
  F_{+} &=& - \frac{1}{2} ( 1 + \cos^{2} \theta) \cos 2\phi \cos 2\psi
            - \cos \theta \sin 2\phi \sin 2\psi \, , \nonumber \\
  F_{\times} &=& \frac{1}{2} ( 1 + \cos^{2} \theta) \cos 2\phi \sin 2\psi
                 - \cos \theta \sin 2\phi \cos 2\psi .
\end{eqnarray}

For a low-mass binary coalescence, the portion of the waveform which is
available to the LIGO and Virgo detectors can be calculated using the
restricted post--Newtonian expansion to a high accuracy
\cite{Blanchet:1995ez,Blanchet:1995,Blanchet:1996pi,Blanchet:2001ax,
Blanchet:2004ek,Damour:1998zb,BuonannoDamour:1999,BuonannoDamour:2000,
Damour:2000zb,Apostolatos:1994, Apostolatos:1995}.  For restricted
waveforms, only the leading order term in the amplitude is used, while
post--Newtonian corrections to the phase are retained.  The waveform
depends upon the masses and spins of the binary components, the
orientation and distance of the binary relative to the detector.  In
this paper, we restrict attention to searches which make use of
waveforms appropriate for binaries without spin.  In Section
\ref{sec:waveform_error} we estimate the effect of using non-spinning
templates in the search for spinning waveforms.

The post--Newtonian expression for the binary inspiral waveform can be
substituted into Eq.~(\ref{eq:ht}) above to obtain an alternative
expression for the gravitational-wave strain at the detector
\cite{findchirp, findchirppaper}.  In this form, the restricted waveform is
expressed in terms of the two phases --- $h_{c}$, the cosine phase, and
$h_{s}$, the sine phase --- of the chirp waveform, and an overall
amplitude and phase factor.  By doing so, there is a clear split between
the overall scaling factors which depend upon the distance, sky location
and orientation of the binary and the mass dependent time evolution of
the waveform.  The waveform is written
\begin{equation}\label{eq:wav}
   h(t) = \frac{1 \mathrm{Mpc}}{\deff} 
  \left[ h_c(t) \cos \Phi + h_s(t) \sin \Phi \right] \, , 
\end{equation}
where the amplitude factor $\deff$ is known as the effective distance to
the binary and $\Phi$ is the coalescence phase as observed at the
detector.  Both the effective distance and coalescence phase are
determined by the location and orientation of the binary system relative
to the detector.  More specifically, the effective distance is defined
as \cite{thorne.k:1987}
\begin{equation}\label{eq:deff}
  \deff = \frac{r}{\sqrt{F_{+}^{2}(1 + \cos^{2}\iota)^{2}/4 +
  F_{\times}^{2} \, \cos^{2} \iota } } \, ,
\end{equation}
where $r$ is the physical distance to the binary, $\iota$ is the
inclination angle of the binary system and $F_{+}$, $F_{\times}$ are
given in (\ref{eq:det_response}).  Similarly, the phase angle $\Phi$ is
dependent upon the sky location, polarization, inclination and also the
coalescence phase $\phi_{o}$ of the binary.

Define the Fourier transform of $h(t)$ by 
\begin{equation} 
\tilde{h}(f) = \int_{-\infty}^{\infty} h(t) e^{-2 \pi i f t} dt \; . 
\end{equation}
The sine and cosine phases of the binary inspiral waveform are dependent
upon the component masses. In the frequency domain, they are obtained
using the stationary phase approximation to the post--Newtonian
expansion \cite{Droz:1999qx}.  In this approximation, $\tilde{h}_{s}(f)
= i \tilde{h}_{c}(f)$, and
\begin{equation}\label{eq:2pn_waveform}
  \tilde{h}_{c}(f) = 
    \mathcal{N} \left[ \frac{G \mchirp}{c^{3}} \right]^{5/6}
                 \Theta(f - f_{\mathrm{max}})
                 f^{-7/6} e^{i \Psi(f;\mchirp,\eta)} \, , 
\end{equation}
where $\mathcal{N}$ is an overall (known) scaling, and we have
introduced the chirp mass $\mchirp = M \eta^{3/5}$, where $M = (m_{1} +
m_{2})$ is the total mass and $\eta = m_{1} m_{2}/ M^{2}$ is the
symmetric mass ratio.  The phase evolution is governed by
$\Psi(f;\mchirp,\eta)$ which also depends upon the masses of the binary
system.  The waveform terminates at a frequency $f_{\mathrm{max}}$.
Physically, the post--Newtonian waveform is expected to break down
around the frequency where the evolution changes from a slow inspiral to
a rapid merger.  A reasonable estimate of this frequency is given by the
innermost stable circular orbit (ISCO) of the Schwarzschild spacetime
with the same total mass $M$, 
\begin{equation}\label{eq:isco}
  f_{\mathrm{isco}} = \frac{2.8 M_{\odot}}{M} 1600 \mathrm{Hz} \, .
\end{equation}

In principle, the waveform observed at the detector for a non-spinning
binary system depends upon eight parameters: the masses of the two
binary components, the physical distance $r$, the sky location and
polarization $(\theta, \phi, \psi)$, the inclination angle $\iota$ and
the coalescence phase $\phi_{o}$.  However, it is clear from
Eq.~(\ref{eq:2pn_waveform}) that the six quantities describing the
location and orientation of the binary affect the strain observed at a
single detector only in the combinations $\deff$ and $\Phi$.
Furthermore, the ability to distinguish a gravitational wave from a
coalescing binary above the background noise is independent of the
coalescence phase at the detector.  Finally, it is only the chirp mass
which affects the amplitude of the waveform (the mass ratio $\eta$ does
affect the phase evolution).  Therefore, the detection efficiency will
depend primarily upon the effective distance $\deff$ and the chirp mass
$\mchirp$.  This observation will be used to greatly simplify the rate
calculation.

\subsection{Inspiral Search Method}
\label{sec:match_filter}

Since the inspiral waveform is well known, the standard matched
filtering technique is used to distinguish signal from noise in a single
detector \cite{wainstein:1962}. Here, we provide a very brief review of
the search implementation, further details are available in
Ref.~\cite{findchirp, findchirppaper}.  The gravitational waveform from
a coalescing binary given in Eq.~(\ref{eq:2pn_waveform}) depends upon
the masses, effective distance and coalescence phase of the binary.  The
two mass dimensions are searched by covering the mass space with a
template bank which guarantees that for any signal there is a good
overlap between the waveform and the closest template \cite{BBCCS:2006}.
As discussed below, the coalescence phase are analytically maximized
over in the matched filtering process and the distance is measured. 

The output of a detector is
\begin{equation}
  s(t) = n(t) + h(t)
\end{equation}
where $n(t)$ is the instrumental noise and $h(t)$ is some, possibly
absent, signal.  The sensitivity of the instrument is characterized by
the (one-sided) power spectrum $S(f)$, given as
\begin{equation}
  \langle \tilde{n}(f) \tilde{n}^\ast(f') \rangle = 
  \half \delta(f-f') S(|f|)
\end{equation}
where the tilde represents the Fourier transform, and the angle brackets
denote the expectation value over the noise.  In order to construct the matched filter,
we introduce an inner product
\begin{equation}
  (a | b) := 4 \, \mathrm{Re} \int_{0}^\infty df 
          \, \frac{\tilde{a}(f)\tilde{b}^\ast(f)}{S(|f|)} \, .
\end{equation}
The sensitivity of the detector to a given signal is characterized by
\begin{equation}
  \sigma^2 = (h_c | h_c) \, ,
\end{equation}
where $\sigma^{2}$ depends upon the noise curve of the instrument as
well as the masses of the binary components (recall that $h_{c}$ is the
waveform at an effective distance of $1$ Mpc).  The SNR is defined as
\begin{equation}\label{eq:snr}
  \rho^2 = \frac{(s | h_c)^2 + (s | h_s)^2}{\sigma^2} \, .
\end{equation}
The analytic maximization over the unknown phase angle has already been
performed by including the outputs from the two orthogonal filters,
$h_{c}$ and $h_{s}$, while the amplitude of the signal determines the
value of the SNR.  

In the absence of a signal, the SNR squared is $\chi^2$ distributed with
two degrees of freedom --- one for each of the filters.  Thus for a
single trial, the probability of obtaining an SNR greater than
$\rho_{\ast}$ is 
\begin{equation}\label{eq:snr_dist}
  P(\rho^{2} > \rho_{\ast}^{2}) = e^{-\rho_{\ast}^2/2} \, .
\end{equation}

If the detector output contains a signal $h(t)$, we characterize its
amplitude by
\begin{equation}\label{eq:rho_h}
  \rho_{h}^{2} := (h | h) = \frac{\sigma^{2}}{\deff^{2}} \, . 
\end{equation}
In this case, the SNR squared is distributed as a non-central $\chi^2$
distribution with two degrees of freedom and a non-centrality parameter
$\rho_{h}^{2}$.  Thus, the expected SNR squared is
$\rho_{h}^{2} + 2$ while the variance is $4(\rho_{h}^{2} + 1)$.  For
SNRs well above unity, the expected SNR is close to $\rho_{h}$.

In the course of a gravitational wave search, we calculate the SNR for
every mass template in the bank and for every time point.  This is used
to construct a list of times and associated mass parameters at which the
SNR exceeds some pre-determined threshold.  These candidate events may
be due to instrumental noise or gravitational waves. 
In searching a
year of data over a wide range of masses, we obtain a background of
events due to noise with a distribution consistent with
(\ref{eq:snr_dist}).  In addition, the data from the detectors contains
non-stationarities which also produce events with high SNR.  The ability
to reduce this background of non-gravitational wave induced events
affects the sensitivity of the search.  There are many techniques
employed to achieve this \cite{LIGOS3S4Tuning}.  The most powerful tool
is a consistency test between detectors --- the arrival time of the
signal should be consistent, within the light travel time between the
sites, and the mass parameters should agree within the search accuracy.
In addition, signal consistency tests, such as the $\chi^{2}$
\cite{Allen:2004} and $r^{2}$ tests \cite{Rodriguez:2007} are
utilized, and an ``effective SNR'', constructed using the SNR and
$\chi^{2}$ information, is used to distinguish signal from noise
\cite{LIGOS3S4all, LIGOS3S4Tuning}.  
After applying these tests, the
typical loudest surviving background events for BNS searches have a
combined SNR $\rho = \sqrt{\rho_{1}^{2} + \rho_{2}^{2}}$ of around $10$.
For higher mass systems, the waveforms spend a shorter time in the
sensitive band.  Consequently, it is more difficult to distinguish them
from noise non-stationarities whence the background extends to higher
SNR.   

\section{Rate calculations for inspiral searches}
\label{sec:upper_limit}

Let us assume that a search for coalescing binaries has been performed
on a stretch of data from gravitational-wave detectors.  We would like
to use the search results to make a statement about the rate of binary
coalescences in the universe.  This can be done by making use of the
loudest event statistic, as described in \cite{loudestGWDAW03,ul}.  The
result will depend upon the astrophysical model for the distribution of
binary coalescences.  To proceed, we make the simple, yet
astrophysically reasonable \cite{Phinney:1991ei}, assumption that binary
coalescences occur only in galaxies and furthermore the rate of binary
coalescence in a given galaxy is directly proportional to the blue light
luminosity of that galaxy.  This assumption is justified by the fact
that both the star formation rate and supernova rate are proportional to
the blue light luminosity.  The result from a search for coalescing
binaries, in the absence of a detection, will be a bound on the rate $R$
of binary inspirals per year per $L_{10} = 10^{10} L_{B,\odot}$, where
$L_{B,\odot}$ is the blue light luminosity of the sun. Recent papers
have suggested that due to the large delay between formation and
coalescence for binaries, this simple assumption will underestimate the
contribution from elliptical galaxies, particularly for BBH rates
\cite{pacheco:2005}.  The formalism presented below can be modified in a
straightforward manner to take this into account.

The loudest event statistic makes use of the fact that no events with an
SNR greater than that of the loudest event $\rhom$ occurred in the
search.  For an inspiral rate $R$, the probability of there being no
gravitational-wave signals with SNR greater than $\rho$ is%
\footnote{The notation $F$ is used to denote ``foreground'' or
gravitational-wave events, in contrast to background $B$ events
associated to instrumental noise.}%
\begin{equation}
  P_{F}(\rho) = e^{ - R \cl(\rho) T } \, .
\end{equation}
where $\cl(\rho)$ is the total luminosity to which the search is
sensitive and $T$ is the duration of the search.  Similarly, the
probability of obtaining zero accidental noise (or background) events
with an SNR greater than $\rho$ is denoted $P_{B}(\rho)$.  Therefore,
the overall probability of obtaining no events with SNR greater than
$\rho$ is
\begin{equation}
  P(\rho | B, R, T) = P_{B}(\rho) e^{ - R \cl(\rho) T } \, .
\end{equation}

Given that no event was observed with SNR greater than that of the
loudest event $\rhom$, a straightforward application of Bayes theorem
leads to the posterior probability distribution for the rate $p(R |
\rhom,T,B )$ which depends upon the loudest event, amount of time
searched and the accidental rate (encoded as `B') \cite{ul}.  In
addition, it will depend upon the prior probability distribution for the
rate, denoted $p(R)$.  In the absence of previous knowledge of the rate,
a uniform prior $p(R) = const.$ is appropriate, while if a previous
search has been performed, the posterior of that search is naturally
used as the prior for the next search.  The expression for the posterior
distribution is 

\begin{equation}\label{eq:rate_posterior}
  p(R | \rhom,T,B ) \propto p(R) 
  \left[ \frac{1 + \Lambda R \, \cl(\rhom) \,  T}{1 + \Lambda}\right]
  e^{- R \, \cl(\rhom) \, T} \, . 
\end{equation}
Here, $\Lambda$ is a measure of the likelihood that the loudest
event is a due to a gravitational wave, rather than from instrumental
noise.  The expression for $\Lambda$ is
\begin{equation}\label{eq:lambda}
  \Lambda =  
  \frac{|\cl^{\prime}(\rhom)|}{P_{B}^{\prime}(\rhom)}
  \left[ \frac{\cl(\rhom)}{P_{B}(\rhom)} \right]^{-1}
  \, ,
\end{equation}
where the derivatives are taken with respect to $\rho$.  The likelihood
depends upon the population through the total luminosity
$\cl(\rhom)$ and its derivative with respect to SNR,
$\cl^{\prime}(\rhom)$.  Similarly it depends upon the background
distribution of noise events through $P_{B}(\rhom)$ and its derivative.
For practical calculational purposes, it is often useful to write 
\begin{equation}\label{eq:nf_and_nb}
  \Lambda = \frac{n_{F}}{n_{B}} 
  \quad \mathrm{where} \quad
  n_{F} = \frac{|\cl^{\prime}(\rhom)|}{\cl(\rhom)}
  \, , \,
  n_{B} = \frac{P_{B}^{\prime}(\rhom)}{P_{B}(\rhom)} \, .
\end{equation}
By doing this, the dependence upon the estimated background is confined
to $n_{B}$ while the quantity $n_{F}$ depends only upon the astrophysical
population model.  

The posterior distribution obtained in Eq.~(\ref{eq:rate_posterior}) can
be used to calculate a Bayesian upper limit on the rate.  The upper
limit $R_{\ast}$ for a given confidence level ($\alpha$) is obtained by
evaluating
\begin{equation}\label{eq:ul_integral}
  \alpha = \int_{0}^{R_{\ast}} p(R | \rhom, T, B) \, .
\end{equation}
Assuming a uniform prior on the rate, the upper limit is given by
\begin{equation}\label{eq:bayesianprob}
  1 - \alpha = e^{-R_{\ast}\, T \, \cl(\rhom)}  
    \left[ 1 + 
    \left( \frac{\Lambda}{1+\Lambda} \right)
    \, R_{\ast}\, T \, \cl(\rhom) \right] \, .
\end{equation}
In the case where the loudest event candidate is most likely due to the
background $\Lambda \rightarrow 0$ and the upper limit becomes $R_{90\%}
= 2.3/[T \, \cl(\rhom)]$.  

In the limit that $\Lambda \rightarrow \infty$, the event is almost
definitely due to a gravitational wave rather than from the noise
background.  In this limit, the probability distribution in
Eq.~(\ref{eq:rate_posterior}) is peaked away from zero, whence it makes
sense to construct a rate interval (as described in \cite{ul}) rather
than a rate upper limit.  This is done by performing the integral in
(\ref{eq:ul_integral}) from $R_{\mathrm{min}}$ to $R_{\mathrm{max}}$.
Alternatively, one could choose to still place an upper limit, in which
case $R_{90\%} = 3.9/[T \, \cl(\rhom)]$.
 
In the remainder of this section, we discuss how the quantities
$\Lambda$ and $\cl$, can be calculated.  In Section
\ref{sec:background}, we describe the estimation of the noise background
and the evaluation of $n_{B}$.  Then, in Section \ref{sec:foreground},
we describe the calculation of the luminosity $\cl$ and consequently
$n_{F}$.  Finally, in Section \ref{sec:the_upper_limit}, we combine
these results to obtain an expression for the upper limit.

\subsection{Estimating the Noise Background} 
\label{sec:background}

In order to calculate an upper limit, we require an estimate of the
background of events due to noise in the detectors.  For perfectly
Gaussian, stationary detectors, this can be calculated theoretically
using the known distribution for the SNR.  However, real detectors
suffer from non-stationarities and time varying sensitivity.  Thus, for
a search involving several detectors, the background is most readily
estimated by time shifting the data from the detectors relative to one
another, and then searching for events which are coincident in time and
mass between the detectors.  If the time shifts are greater than the
light travel time and length of the signal, then the time shifted
coincidences cannot be due to gravitational waves, and are therefore
expected to give a good estimate of the background.  Each time shift
will have a loudest event, which we assume to be drawn from the actual
background distribution for the loudest event, $p_{B}(\rho) :=
P_{B}^{\prime}(\rho)$.  Therefore, by performing a large number of time
shifts, we obtain a good sampling of $p_{B}(\rho)$ from which it is
straightforward to obtain the cumulative distribution $P_B(\rho)$ and
$n_{B} = p_{B}(\rhom)/P_{B}(\rhom)$.

As an example, consider a pair of detectors whose noise output is
Gaussian and stationary.  In this case, the noise background for a
single detector is Poisson distributed, with rate of events louder than
$\rho_{i}$ is given by
\begin{equation}\label{eq:poisson_rate}
  \mu(\rho_{i}) = C e^{-\rho_{i}^{2}/2} \, .
\end{equation}
where $C$ depends upon the number of trials.  With two detectors, the
combined SNR is given by $\rho^2 = \rho_{1}^{2} + \rho_{2}^{2}$.
Furthermore, it is necessary to impose a single
detector threshold on the SNR, denoted $\rho_{T}$.  In this case, the
Poisson rate of background events is
\begin{equation}\label{eq:two_det_rate}
  \mu(\rho) = C( 1 + \rho^{2}/2 - \rho_{T}^{2}) e^{-\rho^{2}/2}
\end{equation}
The constant $C$ can be determined from the expected loudest event.
Given the Poisson rate $\mu(\rho)$, it is straightforward to calculate
the distributions of $P_{B}$, $p_{B}$ and $n_{B}$: 
\begin{equation}\label{eq:poisson_bg}
  P_{B}(\rho) = e^{-\mu(\rho)}
  \, , \, 
  p_{B}(\rho) = |\mu^{\prime}(\rho)| e^{-\mu(\rho)} 
  \, \mathrm{and} \,
  n_{B} = |\mu^{\prime}(\rho)| \, .
\end{equation}
Given the background of (\ref{eq:poisson_rate}), at the expected loudest
event, $n_{B}(\rhom) \approx \rhom$.

\begin{figure}
\includegraphics[width=1.0\linewidth]{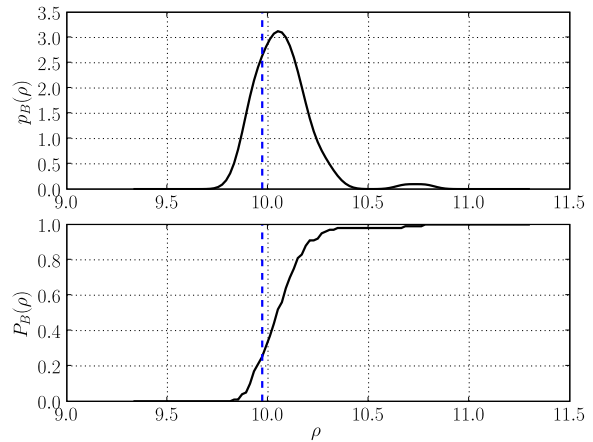}
\caption{\label{fig:slide_dist}
The distribution of the SNR of the loudest event.
These distributions were obtained by generating $100$ time-shift loudest
events from the distributions in Eq.~(\ref{eq:poisson_bg}) with an event
rate $\mu(\rho)$ given by Eq.~(\ref{eq:two_det_rate}), with the
normalization constant chosen so that the expected rate is unity at an
SNR of $10$.  The upper plot shows the probability distribution
$p_{B}(\rho)$, while the lower plot shows the cumulative distribution
$P_{B}(\rho)$.  The distribution is, as expected, peaked close to SNR of
10.  The cumulative probability tends to zero at small SNR --- it is
almost guaranteed that there will be an event louder --- and unity at
large SNR --- it is very unlikely to have an event
louder than this. The dashed line in both plots shows the simulated result
with an SNR of 9.95.} 
\end{figure}

In our example, we choose $\rho_{T} = 5.5$ and select $C$ such that the
expected loudest event has an SNR of $10$, i.e.  $\mu(10) = 1$,
equivalent to an SNR of about 7 in each detector.  For this search, we
simulate $100$ time shifts and obtain the loudest event for each.  The
distributions are plotted in figure \ref{fig:slide_dist}.  The features
in these plots are due to the finite number of time-shifts performed,
which lead to uncertainties in the reconstruction of the distributions.
In addition, we simulate the output of the search, and obtain a loudest
event with $\rhom = 9.95$ which yields values of $p_{B} = 2.7$, $P_{B} =
0.25$ and $n_{B} = 10.9$.
 
\subsection{Calculating the Foreground}
\label{sec:foreground}

The upper limit derived from a search depends upon the total luminosity
$\cl$ to which a search is sensitive.  This must be evaluated at the SNR
of the loudest event $\cl(\rhom)$.  In section \ref{sec:inspiral}, we
have shown that the strength of the gravitational-wave signal at a
detector depends primarily upon the effective distance $\deff$ and chirp
mass $\mchirp$ of the signal.  In addition, the measured SNR of signal
with a given $\deff$ and $\mchirp$ will fluctuate depending upon the
noise in the detectors at the time.  Thus, for each value of $\deff$ and
$\mchirp$, we can calculate the probability that the signal is observed
with an SNR greater than $\rho$.  This quantity is known as the
efficiency, $\epsilon(\rho, \deff,\mchirp)$.  Since the sensitivity of a
detector varies over the course of a search, the efficiency is measured
by performing Monte Carlo simulations.  For a search involving several
detectors, the efficiency will depend upon the effective distance to the
source from all detectors, which we denote in bold as $\mathbf{\deff}$.
The efficiency calculation is discussed in Section \ref{sec:efficiency}.  

To calculate the total luminosity, we also require a measure of the blue
light luminosity $L_{B}(\deff,\mchirp)$ as a function of $\deff$ and
$\mchirp$.  This is calculated from the known luminosity density in the
universe.  For the initial LIGO detectors --- with sensitivity to binary
neutron star coalescences up to 40 Mpc --- it is necessary to take into
account the inhomogeneity of the local universe.  This requires the
construction of a catalog of nearby galaxies (see
Ref.~\cite{LIGOS3S4Galaxies} for details on how this is constructed).
Armed with a galaxy catalog and a mass distribution for the binaries,
the method of calculating $L_{B}$ is described in Section
\ref{sec:astro}.  

Finally, given the efficiency and luminosity functions, the cumulative
luminosity is given by 
\begin{equation}
  \cl(\rho) = \int d\mathbf{\deff}\, d\mchirp \, 
  \epsilon(\rho, \mathbf{\deff},\mchirp) \, 
  L_{B}(\mathbf{\deff},\mchirp) \, .
\end{equation}
Details of this calculation are given in Section \ref{sec:cumlum}.

\subsubsection{Efficiency}
\label{sec:efficiency}

The efficiency of a search is evaluated by adding simulated inspiral
signals to the data stream and evaluating the probability that signals
with a given set of parameters are detected with SNR greater than
$\rhom$.  By performing a host of injections, it is possible to evaluate
the efficiency as a function of both the chirp mass and effective
distance.  

\begin{figure}
\includegraphics[width=1.0\linewidth]{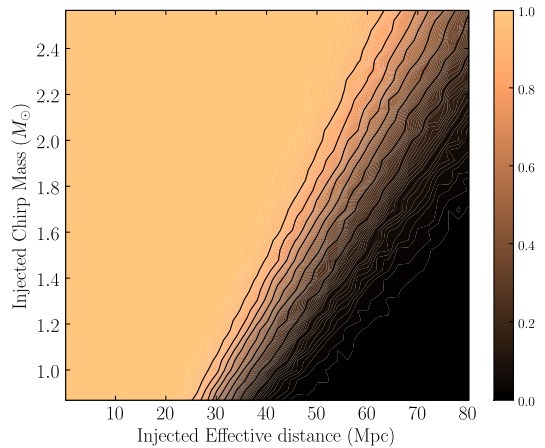}
\caption{\label{fig:sim_efficiency}
The efficiency of the initial LIGO detectors to coalescing binaries, as
a function of the effective distance and chirp mass of the source.  The
efficiency curve is evaluated at an SNR of $7$ in the
detector.}
\end{figure}

Figure \ref{fig:sim_efficiency} shows the simulated efficiency of
the initial LIGO detectors, at an SNR of 7, as a function of the binary's
effective distance and chirp mass.  The distance to which events are
detectable increases with the chirp mass.  This follows immediately from
Eq.~(\ref{eq:2pn_waveform}) for the inspiral waveform, from which we see
that the amplitude is proportional to $\mchirp^{5/6}/\deff$.  For
binaries with a total mass less than $10$ $M_{\odot}$, the inspiral
stage of the evolution will sweep right across the sensitive band of the
detector.  Therefore, for low-mass signals, the detectability of a
signal at a given detector will be dependent only on the chirp distance
$\dchirp$ of the binary:
\begin{equation}\label{eq:dchirp}
  \dchirp := \deff  
  \left(\frac{\mchirp_{1.4}}{\mchirp}\right)^{5/6} \, .
\end{equation}
For higher mass signals, the coalescence will occur within the sensitive
band of the detector, as can be seen from Eq.~(\ref{eq:isco}).  In this
case, there is no simple relationship between the efficiency, chirp mass
and distance, and both mass and effective distance must be taken into
account when calculating the efficiency.

\begin{figure}
\includegraphics[width=1.0\linewidth]{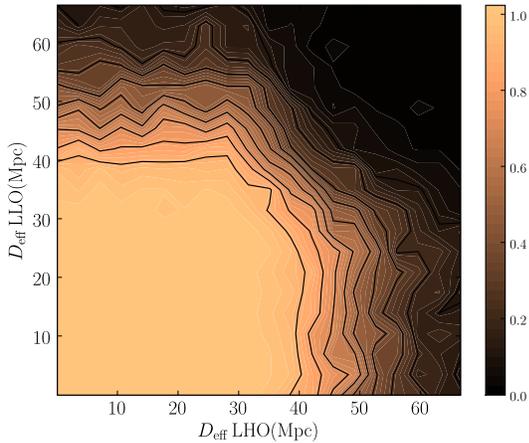}
\caption{\label{fig:2d_efficiency}
The detection efficiency as a function of the effective distance for two
detectors.  The efficiency at small distances is unity, while at large
distances it is zero.  The transition is governed by two effects.  At
large distances in one detector, the efficiency is limited by the single
detector threshold of $5.5$.  At large distances in both instruments,
the efficiency is determined by the combined SNR threshold of 10.  The
variations in the contours are due to the fact that a finite number of
injections --- 100 in each bin --- were used when generating the
efficiency curve.}
\end{figure}

Let us now generalize to the case of several detectors.  A given binary
coalescence will have a different effective distance in each detector,
and therefore the efficiency will be a function of the effective
distance in each detector. In figure \ref{fig:2d_efficiency} we plot the
efficiency against effective distance in the H1 and L1 detectors, for a
$1.4$--$1.4 \, M_{\odot}$ binary.  At small effective distance in both
detectors, the efficiency is unity, while at large distances it goes to
zero.  The shape of the constant efficiency contours depends upon two
factors:  the single instrument threshold and the combined threshold.
At large effective distance in one instrument, but small in the other,
the single instrument threshold limits the efficiency.  For comparable
distances, the efficiency is determined by the combined SNR.

\subsubsection{Astrophysical Model}
\label{sec:astro}

In order to calculate the total luminosity to which a search is
sensitive, we require the luminosity density as a function of the
effective distance and mass.  This involves combining a model of the
location and luminosity of galaxies with the antenna response functions
of the detectors, given in Eq.~(\ref{eq:det_response}).

Let us consider a source from a given galaxy.  The effective distance to
the source depends upon the physical distance, and four sky angles ---
the sky location relative to the detector, the inclination and
polarization angles.  Equivalently, the location of the source
can be described by $\lamvec = (D, \alpha, \delta, \iota, \psi, t)$,
where the right ascension $\alpha$, declination $\delta$ and sidereal
time $t$ serve to specify the source sky location.  For sources from a
particular galaxy, the distance, right ascension and declination of the
galaxy $(D_{i}, \alpha_{i}, \delta_{i})$ are known.  Then, assuming that
binary coalescences are uniformly distributed over the sidereal day,
inclination and polarization angles, we obtain a distribution for the
observed effective distances of sources from a given galaxy 
\begin{eqnarray}
  p_{i}(\deff) &=& \int d\lamvec \, p(t) \, p(\iota) \, p(\psi) \, 
  \delta(D - D_{i}) \,  \delta(\alpha - \alpha_{i}) \nonumber \\
  && \qquad \delta(\delta -\delta_{i}) \,\delta(\deff - \deff(\lamvec)) \, ,
\end{eqnarray}
where $p(t) = 1/(\mbox{sidereal day})$, $p(\psi) = 1/2\pi$ and $p(\iota)
= \sin(\iota)/2$.  In practice, this distribution is most easily
obtained by simulating many signals, at random times and orientations, 
from a given galaxy.

In Ref.~\cite{LIGOS3S4Galaxies}, the compact binary coalescence galaxy
(CBCG) catalog has been compiled to a distance of 100 Mpc.  For galaxies
in this catalog the sky location and distance to the galaxy are known.
In addition, the apparent magnitude $m_{B,i}$ in blue light of the
galaxy is measured.  The luminosity of the galaxy is obtained from its
distance $D_{i}$ and apparent magnitude as
\begin{equation}\label{eq:galaxy_luminosity}
  \frac{L_{B,i}}{L_{B,\odot}} = 
  \left( \frac{D_{i}}{10 \mathrm{pc}} \right)^2 
  10^{(M_{B, \odot} - m_{B,i})/2.5} \, , 
\end{equation}
where $L_{B,\odot} = 2.16 \times 10^{33} \mathrm{ergs/s}$ is the blue
solar luminosity, and $M_{B, \odot}$ is the (absolute) blue solar
magnitude \cite{Binney-Tremaine}.

\begin{figure}
\includegraphics[width=1.0\linewidth]{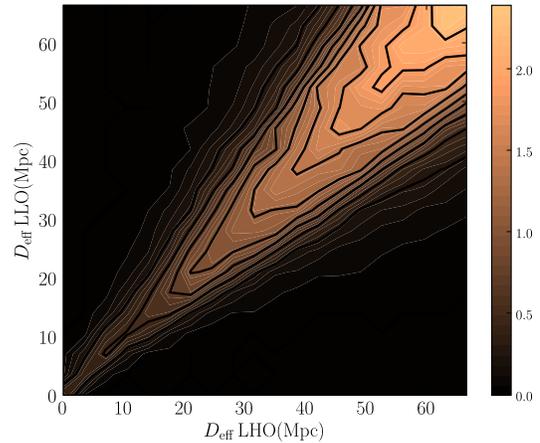}
\caption{\label{fig:luminosity} The luminosity distribution in the
nearby universe as a function of the effective distance in the Hanford
and Livingston detectors.  The distribution is obtained by assuming that
the binary coalescences in a given galaxy are dependent upon the blue
light luminosity of that galaxy and are uniformly distributed in
sidereal time, inclination and polarization.  The color bar indicates
the available blue light luminosity $(L_{10}/\mathrm{Mpc}^{2})$.  The
number increases with distance in both detectors and is greatest on the
diagonal.  The off diagonal spread in luminosity is due to the different
alignments of the Hanford and Livingston detectors.}
\end{figure}

Given the distribution of effective distances for each galaxy, it is
straightforward to obtain the total luminosity as a function of
effective distance by summing over all galaxies:
\begin{equation}
  L_{B}(\deff) = \sum_{i} L_{B,i} \, p_{i}(\deff) \, .
\end{equation}
As before, this can be easily generalized this to a distribution of
luminosity as a function of effective distance for several detectors.
In Figure \ref{fig:luminosity} we make use of the galaxy catalog of
\cite{LIGOS3S4Galaxies} to plot the luminosity as a function of the
effective distance at the Hanford and Livingston detectors.

Finally, we must include the mass distribution $p(\mchirp)$.  The
available luminosity is expressed as a function of effective distance
and chirp mass as
\begin{equation}\label{eq:luminosity}
  L_{B}(\deff,\mchirp) = \left(\sum_{i} L_{B,i} \, p_{i}(\deff) \right) \, 
  p(\mchirp) \, .
\end{equation}

For neutron star binaries, the mass distribution can be taken from
observations \cite{Thorsett:1998uc} or, alternatively, from population
synthesis \cite{Belczynski:2002}.  For binaries containing at least one
black hole, the lack of observations leads to greater uncertainty in the
mass distribution, whence it is more natural to calculate a mass
dependent rate limit.

\subsubsection{Calculating the cumulative luminosity}
\label{sec:cumlum}

The cumulative luminosity available to a search is obtained by
multiplying the luminosity distribution in Eq.~(\ref{eq:luminosity})
against the efficiency $\epsilon(\rho,\deff,\mchirp)$ and integrating over
the mass and effective distance:
\begin{equation}\label{eq:cum_lum}
  \cl(\rho) = 
  \int d \mathbf{\deff} \, d \mchirp \, 
  \epsilon(\rho, \mathbf{\deff}, \mchirp) \,
  L_{B}(\mathbf{\deff}) \, p(\mchirp)\, .
\end{equation}
For a given mass, this reduces to multiplying the data from Figures
\ref{fig:2d_efficiency} and \ref{fig:luminosity} and integrating over
the effective distances.  This must then be integrated over of the chirp
mass to obtain the cumulative luminosity.  For low-mass systems, the
calculation is greatly simplified by recalling that the efficiency
depends only upon the chirp distance (\ref{eq:dchirp}) at a given site.

In a similar manner, the derivative of the cumulative luminosity
$\cl^{\prime}(\rhom)$ can be calculated.  This is done by calculating
the rate of change of efficiency as a function of SNR, evaluated at
$\rhom$.  With this information, we can calculate:
\begin{equation}
  n_{F} = \frac{|\cl^{\prime}(\rhom)|}
  { \cl(\rhom) } \, .
\end{equation}

In cases where the mass distribution is not known, the luminosity can be
expressed as a function of the mass:
\begin{equation}\label{eq:cum_lum_vs_mass}
  \cl(\rho, \mchirp) = 
  \int d \mathbf{\deff} \, \epsilon(\rho, \mathbf{\deff},\mchirp)
  \, L_{B}(\mathbf{\deff}) \, .
\end{equation}
Then, the rate limit is naturally expressed as a function of the
mass.

We can compute the $\cl$ for the example with a loudest event SNR of
$10$.  To simplify the calculation, assume that the mass of all binary
components is $1.4 M_{\odot}$.  Then, the total luminosity and its
derivative are
\begin{equation}
  \cl(\rhom) = 540 \, L_{10} \quad \mathrm{and} \quad
  \cl^{\prime}(\rhom) = 120 \,  L_{10} \, .
\end{equation}
This gives a value of $n_{F} = 0.22$.  For comparison, we can calculate
the expected value for a uniformly distributed population.  In this
case, $\cl \propto \rho^{-3}$, whence $n_{F} \approx 3/\rhom$, which
is similar to the calculated value.  The difference is due to the fact
that at the distances under consideration, the non-uniformity of the
luminosity distribution is important.

\subsection{Obtaining an Upper Limit}
\label{sec:the_upper_limit}

In the preceding sections, we have described all of the pieces which
are required in calculating an upper limit from a search for
gravitational waves from binary coalescence.  The calculation of the
cumulative luminosity $\cl$ depends upon the efficiency of the search,
and the astrophysical distribution of signals.  We have argued that in
the non-spinning case, these are both described as functions of
effective distance and chirp mass.  The likelihood $\Lambda$
depends upon the foreground and background distributions, encoded in
$n_{F}$ and $n_{B}$.  The first of these, $n_{F}$ is obtained from the
cumulative luminosity and its derivative, while $n_{B}$ is determined
from analysis of events arising in analysis of time shifts of the data. 

The formula for the upper limit was given in
Eq.~(\ref{eq:bayesianprob}).  Taking the prior distribution $p(R)$ of
the rate to be uniform, we obtain a Bayesian upper limit with confidence
$\alpha$ as

\begin{equation}\label{eq:ul}
  1 - \alpha = e^{-R\, T \, \cl(\rhom)}  
    \left[ 1 + 
    \left( \frac{\Lambda}{1+\Lambda} \right)
    \, R\, T \, \cl(\rhom) \right] \, .
\end{equation}
For a representative loudest event, we obtain $n_{B} \approx \rhom$
and $n_{F} \approx 3/\rhom$.  Therefore, $\Lambda \sim 3/\rhom^{2}$.
For any realistic search, with a loudest event of SNR around $10$, this
implies that $\Lambda \ll 1$, and therefore the loudest event is most
likely background.  To obtain a mass dependent upper limit, this
analysis is repeated for different values of $\mchirp$ making use of the
mass dependent luminosity function $\cl(\rhom,\mchirp)$ to obtain a rate
limit $R(\mchirp)$.  This method neglects the fact that the distribution
of background events is also mass dependent by using the same loudest
event for all masses.  

Plugging in the values obtained in the previous sections, we have
$\Lambda = 0.22/10.9 = 0.020$, and $\cl(\rhom) = 540 \, L_{10}$.  Assuming
a year of analysis time, the limit would be
\begin{equation}\label{eq:example_ul}
  R_{90\%} = \frac{2.35}{\cl(\rhom) \, T} 
  = 4.3 \times 10^{-3} L_{10}^{-1} \, yr^{-1} \, .
\end{equation}
This gives a reasonable estimate of the expected BNS upper limit in the
absence of a detection in the LIGO S5 science run.

How does this compare with astrophysical predictions?  The rate of
galactic binary neutron star mergers is $8.30^{+20.91}_{-6.61} \times
10^{-5} yr^{-1}$, at 95\% confidence \cite{Kalogera:2004nt}.  Assuming
the rate is indeed a function of the blue light luminosity alone, and
using $1.7 \, L_{10}$ as the blue light luminosity of the Milky Way
\cite{Kalogera:2004nt}, gives an optimistic rate of $1.7 \times 10^{-4}
L_{10}^{-1} \, yr^{-1} $, which is a factor of 25 lower than the expected
upper limit from the analysis of one year of data from initial LIGO.  

\section{Systematic Errors}
\label{sec:systematics}

In the previous section, we have described a method for calculating an
astrophysical upper limit or interval for the rate of binary
coalescences from the results of a gravitational-wave search.  The
probability distribution for the rate is dependent on four quantities:
the prior distribution $p(R)$, the cumulative luminosity $\cl$, the
likelihood $\Lambda$ of the loudest event being a signal, and the
analysis time $T$.  Of these quantities, only the analysis time can be
unambiguously measured.  The choice of prior distribution $p(R)$ will
affect the posterior distribution.  However, we take the prior as an
input to the analysis and do not consider uncertainties associated to
the choice of prior.  There are numerous systematic errors which will
affect the measured values of both the luminosity and likelihood.  These
systematic effects can be broadly split into five categories:

\begin{itemize}

\item Imprecise knowledge of the astrophysical distributions of the mass
and distance of binaries.

\item Differences between the physical signal and the non-spinning,
restricted post--Newtonian waveforms.

\item Statistical fluctuations in the measured efficiency.

\item Uncertainties in instrumental calibration.

\item Errors in the calculated likelihood $\Lambda$, arising from the
above uncertainties and errors in the background estimation.

\end{itemize}

In this section, we describe the various sources of uncertainty and
analyze their effect.  Additionally, we perform a marginalization over
these uncertainties to produce the rate distribution.

\subsection{Uncertainties in population model}
\label{sec:popuncertainties}

The cumulative luminosity of a search will depend critically upon the
astrophysical model used.  In particular, the luminosity distribution is
sensitive to both the location and luminosity of galaxies within the
reach of the search.  In addition, since the amplitude and frequency
range of gravitational waves emitted by a binary coalescence is mass
dependent, the cumulative luminosity will also be dependent upon the
astrophysical mass distribution.  In Section \ref{sec:galaxy_dist}, we
discuss the systematics associated to uncertainties in galaxy
distribution, while in \ref{sec:mass_dist} we investigate the effect of
changing the mass population.

\subsubsection{Galaxy Distribution}
\label{sec:galaxy_dist}

The sky position of nearby galaxies is known accurately enough that
errors in the right ascension and declination will not affect the
cumulative luminosity.  However, the distances and luminosity of
galaxies are not so well known, whence these uncertainties must be taken
into account when calculating the total luminosity.  Indeed, the
luminosity of a galaxy is not directly measurable, instead it is
calculated from the apparent blue magnitude $m_{B,i}$ and distance
$D_{i}$ using (\ref{eq:galaxy_luminosity}).  The uncertainties in
distances vary from less than $10\%$, from observations of Cepheids in
nearby galaxies observed with the Hubble Space Telescope, to
uncertainties up to $30\%$ for more distant galaxies.  Additionally,
there are uncertainties in the apparent magnitude of galaxies which vary
from $\Delta m_{B,i} = 0.3$ to $\Delta m_{B,i} = 0.38$
\cite{LIGOS3S4Galaxies}.

We begin by considering an error in distance of $\Delta D_{i}$ to a
galaxy at distance $D_{i}$.  The change in distance of the galaxy will
lead to a linear change in the effective distance of all sources from
that galaxy.  More precisely, changing the distance from $D_{i}$ 
to $(D_{i} + \Delta D_{i})$ will change to distribution of effective
distances from $p_{i}(\deff)$ to $p_{\Delta, i}(\deff)$ where
\begin{equation}\label{eq:delta_deff_dist}
  {p}_{\Delta, i}[\deff (1 + \Delta D_{i}/ D_{i})] = p_{i}(\deff) \, .
\end{equation}
Thus, the average effective distance to a source will increase by $(1 +
\Delta D_{i}/D_{i})$.  This will not have any effect for close (or
very distant) galaxies where the efficiency to sources from that galaxy
is unity (or zero).  However, for galaxies at distances where the
efficiency is rapidly changing, this can be a significant effect,
reducing the efficiency when the distance to the galaxy is increased.

Since the luminosity of a galaxy is inferred from its measured distance
and apparent magnitude, a change in the distance will also affect the
calculated luminosity.  It follows directly from
(\ref{eq:galaxy_luminosity}) that a change in distance of $\Delta
D_{i}$, leaving the apparent magnitude unchanged, will yield a change in
luminosity $\Delta L_{B,i}$ given by
\begin{equation}\label{eq:luminosity_distance}
  \frac{L_{B,i} + \Delta L_{B,i}}{L_{B,i}} = 
  \left( \frac{D_{i} + \Delta D_{i}}{D_{i}} \right)^2
\end{equation}
Thus, the inferred luminosity will increase if the distance to the
galaxy increases. This effect will be significant for all galaxies to
which the search has non-zero efficiency.  

It is also straightforward to calculate the effect of errors in the
reported apparent magnitude.  We have from
Eq.~(\ref{eq:galaxy_luminosity})
\begin{equation}\label{luminosity_magnitude}
  \frac{L_{B,i} + \Delta L_{B,i}}{L_{B,i}} = 10^{(\Delta m_{B,i}/2.5)} \, .
\end{equation}
Therefore, an increase (decrease) in magnitude will lead to an increase
(decrease) in $\cl$.  The reported errors in the CBCG catalog are
$\Delta m_{B,i}$ between $0.3$ and $0.38$, giving corresponding
uncertainties in the luminosity of $32\%$ and $42\%$ respectively.

In order to estimate the uncertainty in the luminosity $\cl$, it is
necessary to vary the distance and magnitudes of all galaxies for which
the search has non-zero sensitivity.  We expect that the errors are
independent for different galaxies, although it is certainly conceivable
that there is an overall systematic for a given galaxy catalog.
However, to be conservative, we calculate the errors by moving
\textit{all} galaxies either closer or further.  As with the distance
error, we take the conservative error by increasing or decreasing the
magnitude of all galaxies together.

At large distances, the luminosity distribution can be taken as
homogeneous and isotropic.  In this case, the blue luminosity density is
calculated directly from observations as $L_{B} = (1.98 \pm 0.16) \times
10^{-2} L_{10}/ \mathrm{Mpc}$ \cite{LIGOS3S4Galaxies}.  Thus, at
distances greater than 40 Mpc, the uncertainty in luminosity can be
calculated directly.

Recent papers have suggested that due to the large coalescence times for
binary inspiral, a significant fraction of them might occur in
elliptical galaxies where the star formation rate is low.  The general
loudest event formalism presented in Ref.~\cite{ul} can be used in this
case, and one might envision introducing two unknown rate parameters
$R_{B}$ which depends upon the blue light luminosity discussed above and
$R_{E}$ which is a second contribution due to elliptical galaxies.
Then, the rate would depend upon two parameters, and a given search
could be used to place a confidence bound in the two dimensional space
spanned by them.  If the corrections from including elliptical galaxies
prove to be significant, this effect will be folded into future rate
calculations.

\subsubsection{Binary mass distribution}  
\label{sec:mass_dist}

There is significant uncertainty in the mass distribution of coalescing
binaries.  Several binary neutron star systems, and significant numbers
of single neutron stars, have been observed as pulsars, allowing us to
place some restrictions on the mass distribution.  However, the mass
distributions presented in (for example) \cite{Belczynski:2002} are
produced using large scale simulations which must assume an equation of
state for the nuclear material.  For stellar mass black hole binaries,
there is little restriction on the mass distribution.  These
uncertainties lead us to place mass dependent upper limits.  However, it
is still instructive to look at the sensitivity of our binary neutron
star search for various mass distributions.

As discussed above, the distance to which a source can be observed is
dependent upon its chirp mass.  To leading order, the amplitude of the
emitted gravitational radiation, and hence the distance to which the
source can be observed, is proportional to $\mchirp^{5/6}$.  The
astrophysical mass distribution of neutron stars can have a significant
effect upon the distance to which sources are observable.  For a
$1.4-1.4 M_{\odot}$ solar mass binary ($\mchirp = 1.22 M_{\odot}$),
the 50\% efficiency point for initial LIGO at SNR 7 occurs at 40 Mpc,
whereas for a $3.0-3.0 M_{\odot}$ binary that is increased to 75 Mpc.
So, an astrophysical population containing higher mass binaries will
increase our range.  As an example, let us consider two mass
distributions:
\begin{itemize} 

\item The distribution of observed masses of radio pulsars
\cite{Thorsett:1998uc}, namely a Gaussian distribution peaked at $1.35
M_{\odot}$, with a width of $0.04 M_{\odot}$.

\item The distribution from Ref.~\cite{Belczynski:2002} obtained from
population synthesis models.  This is the mass distribution which was
used in obtaining the LIGO S1 and S2 results given in
Refs.~\cite{LIGOS1iul, LIGOS2iul}.

\end{itemize}

\begin{figure}
\includegraphics[width=1.0\linewidth]{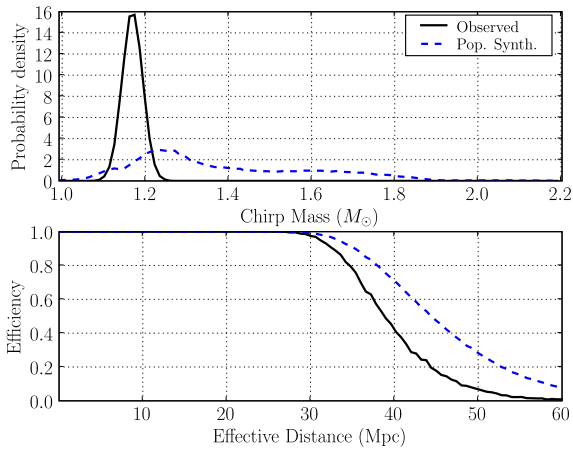}
\caption{\label{fig:mass_distributions} 
Effect of the mass distribution on the search efficiency.  The upper
plot shows two different binary neutron star mass
populations: one taken directly from observed Neutron star
masses, which are fitted to a Gaussian mass with a peak at $1.35 M_{\odot}$
and width of $0.04 M_{\odot}$; the other from population synthesis
models in Ref.~\cite{Belczynski:2002}.  The lower plot shows the
efficiency as a function of effective distance for the initial LIGO
detector at SNR $7$, where mass has been folded in.}
\end{figure}

\noindent Figure \ref{fig:mass_distributions} shows the distribution of
chirp mass for the two different cases.  In both cases, the peak is at
around the same chirp mass of $1.2 M_{\odot}$, corresponding to binaries
with component mass around $1.4 M_{\odot}$.  However, there is a
significant fraction of higher mass neutron stars in the population
synthesis distribution.  

Given a mass distribution, we can integrate the efficiency over the mass
to obtain efficiency as a function of effective distance alone,
\begin{equation}
  \epsilon(\deff) = \int d \mchirp \, p(\mchirp) \,
  \epsilon(\deff,\mchirp) \, .
\end{equation}
This curve is also plotted in Figure \ref{fig:mass_distributions} for
the two mass distributions of interest.  Varying the mass distribution
has a 20\% effect on the distance at which the efficiency reaches 50\%.
Since the choice of mass distribution can have a significant effect on
the efficiency, in \cite{LIGOS3S4all} the upper limits from LIGO
searches are reported for a stated sample distribution.  The mass
uncertainty is not folded into the systematic errors on the upper limit.

\subsection{Uncertainties in the Waveform}   
\label{sec:waveform_error}

In several places, we have assumed that both the physical waveforms and
the search templates are described by the post--Newtonian approximation
and we have ignored the effects of spin.  The methods presented here are
not tied to the particular templates used and extend immediately to
searches using other types of waveform, or even to untemplated,
excess-power type analyses \cite{Anderson:2000yy}.  The only requirement
is that it is possible to meaningfully associate an SNR to both noise
and simulated signals, and hence obtain a loudest event.  Throughout, we
have also modeled the true astrophysical waveforms as those obtained
from the restricted, non-spinning, post--Newtonian calculation.  This can
have a significant effect on the interpretation of the result.

In order to explore how the uncertainty in the simulated waveforms will
affect our result, it is useful to introduce the notion of match.  For a
given set of parameters (in particular the component masses and
effective distance), we denote the true waveform by $h_{t}$ and
those used in simulations as $h_{s}$.  The difference between the true
and simulated waveforms is encoded in the match M defined as 
\begin{equation}\label{eq:match}
  M = \frac{(h_{t}|h_{s})}{\sqrt{(h_{t}|h_{t})(h_{s}|h_{s})}} \, .
\end{equation}
If the waveforms agree perfectly, the match will be unity, while in all
other cases it will be less than one.

Differences between the post--Newtonian approximation and the true
waveforms have been examined in some detail.  In Refs.~\cite{Droz:1997,
Droz:1998}, this has been done by comparing the post--Newtonian
waveforms to those obtained from black hole perturbation theory.  The
results indicate approximately a $10\%$ loss of SNR (i.e. a match of
$90\%$) due to inaccurate modelling of the waveforms for low-mass
systems, with the effect becoming more pronounced for higher mass
ratios.  In Ref.~\cite{Canitrot:2001hc} a similar result was obtained by
comparing waveforms at different post--Newtonian order. Recent
breakthroughs in numerical simulations of black hole and neutron star
coalescences will allow for further exploration of this issue.  

In order to fully address the issue of spin, it will be necessary to
perform substantial Monte--Carlo simulations of spinning waveforms.  For
the time being,  we rely on estimates provided by Apostolatos
\cite{Apostolatos:1995} in which he shows that less than 10\% of all
spin-orientations and parameters consistent with binary neutron stars
provide a loss of SNR greater than $5\%$.  

In BNS searches, where the same waveforms are used as templates and
simulated signals, it is straightforward to calculate the effect of any
mismatch.  In this case, the mismatch between the astrophysical waveform
and the post--Newtonian approximation will lead to an over-estimation of
the observed SNR from a given binary coalescence.  Specifically, if
$\rho_{t}$ and $\rho_{s}$ are the SNRs associated to the true signal and
simulation respectively, then
\begin{equation}\label{eq:true_snr}
  \rho_{t} = \rho_{s} M \sqrt{\frac{(h_{t} | h_{t})}{(h_{s} | h_{s} ) }}
  \, .
\end{equation}
Therefore, the SNR associated to a true signal is reduced by a factor
$M$ from what is observed in a simulation.  

There is an important difference between the waveform uncertainties and
the other systematic errors discussed in this section.  In the
simulation, we are using the same waveform for injection and detection.
In reality, the true astrophysical waveforms will not match precisely
the detection family.  This will lead to a decrease in the overlap
between the astrophysical and detection families.  It is not possible
for this to lead to an increase as the match cannot be greater than
unity.  Thus, the waveform errors can only serve to decrease the
cumulative luminosity available to a search.  However, in cases where
the simulated signals and templates do not match precisely, it is
possible that the ``true'' waveforms will have a better match with the
templates than the simulations do.  Then, errors in the waveform may 
cause us to underestimate the efficiency of the search.   

Returning to Eq.~(\ref{eq:true_snr}), we see that the overall
normalization of the waveform will also affect the SNR.  Generally, it
is assumed that the amplitude of the simulated and astrophysical
waveforms are in good agreement, namely $(h_{s}|h_{s}) \approx
(h_{t}|h_{t})$.  In Refs.~\cite{VanDenBroeck:2006qi,
VanDenBroeck:2006qu}, the authors consider the effects of higher order
post--Newtonian corrections to the amplitude.  These are particularly
important for higher masses, especially when the ratio of the component
masses is large.  Furthermore, in \cite{VanDenBroeck:2006qi} it has been
noted that the inclusion of amplitude corrections actually decreases the
overall amplitude of the signal.  Thus, even though neglecting amplitude
corrections may not significantly affect the detection ability of a
search, it can still have an effect on the interpretation of results.
This effect is not important for BNS systems, but does become important
in higher mass, asymmetric systems.

\subsection{Uncertainties in the instrumental calibration}
\label{sec:respuncertaninties}

When performing simulations of the gravitational-wave signal from a
coalescing binary, the SNR $\rho_{s}$ associated with a simulated
signal differs from the SNR $\rho_{t}$ that would be associated with
a real signal with the same parameters due to uncertainties in the
instrumental calibration.  

In calculating the efficiency of a search, simulated events are added in
software to the data after it has been recorded.  Therefore, any
uncertainty in instrumental calibration will not affect the software
injections in the way it will a real signal.  To quantify this effect,
we focus on the differences between a true signal with given masses and
effective distance $\deff$ and a simulated signal with the same
parameters.  To simplify matters, we assume that the waveform exactly
matches one of the search templates (i.e. ignore the waveform
uncertainty discussed above), in which case
\begin{equation}\label{eq:swstrain}
  \tilde{s}(f) = \frac{e^{i \phi_0}}{\deff} \tilde{h}_{c}(f) +
  \tilde{n}(f) \, ,
\end{equation}
where $\tilde{h}_{c}(f)$ is the waveform introduced in
(\ref{eq:2pn_waveform}), $\phi_{0}$ is an arbitrary phase, and $n(f)$ is
the detector noise.  Then, by substitution into Eq.~(\ref{eq:snr}), it
is straightforward to verify that $ \langle \rho^{2} \rangle =
\rho_{h}^{2} + 2$ as expected.  

The output of a gravitational-wave detector is not the gravitational
wave strain, it is an uncalibrated signal $v(t)$.  This output is
related to the strain by
\begin{equation}
  \tilde{s}(f) = R_{t}(f) \tilde{v}(f)
\end{equation}
where $R_{t}(f)$ is the true response function of the instrument.  The
process of calibrating the data involves the construction of a response
function $R(f)$.  Due to calibration uncertainties, the reconstructed
response will differ from the true response.  These calibration
uncertainties can, to some degree, be independently tested by performing
``hardware injections'' into the detectors \cite{Brown:2004}.  To
calculate the effect of calibration uncertainties, let us follow
\cite{Allen:1996} and write the measured response as
\begin{equation}\label{eq:response}
  R(f) = r(f) e^{i \theta(f)}
\end{equation}
and the true response as 
\begin{equation}\label{eq:true_response}
  R_{t}(f) = [r(f) + \delta r(f)] e^{i [\theta(f) + \delta \theta(f)]}
  \,  .
\end{equation}
where $\delta r$ and $\delta \theta$ are the amplitude and phase parts
of the calibration error which are assumed to be small.  

In the event that a gravitational wave with strain $h(t)$ impinges on
the detector, the calibrated strain reconstructed from the output of the
detector will be given by:%
\footnote{For simplicity, we ignore contributions from the noise in the
following.}
\begin{equation}\label{sigstrain} \tilde{s}_{t}(f) =
\frac{R(f)}{R_{t}(f)} \frac{e^{i \phi_0} \tilde{h_{c}}(f)}{\deff} \, .
\end{equation}
The SNR for an event observed in the detector can be calculated by
substituting Eq.~(\ref{sigstrain}) into the expression (\ref{eq:snr})
for the SNR.  Then, making use of the expressions (\ref{eq:response})
and (\ref{eq:true_response}) for the response functions, and expanding in
powers of $\delta r/r$ and $\delta \theta$ we obtain
\begin{eqnarray}\label{eq:measured_snr}
  \rho_{t}^{2} &=& \rho_{h}^{2} 
  - \frac{\rho_{h}^{2}}{\sigma^{2}} \left[ 4 \int_0^{\infty} df
    \frac{|\tilde{h}_{c}(f)|^2}{S_n(f)} 
    \left( \frac{2 \delta r}{r} + (\delta \theta)^2 \right) \right]
    \nonumber \\
  && \quad + \frac{\rho_{h}^{2}}{\sigma^4} \left(4 \int_0^{\infty} df
    \frac{|\tilde{h}_{c}(f)|^2}{S_n(f)} \delta \theta \right)^2 \, .
  \end{eqnarray}
Thus, the error is linear in $\delta r$ but quadratic in $\delta
\theta$.  Furthermore, it follows directly from the Schwarz inequality
that 
\begin{equation}
  |(\tilde{h}_{c} \delta \theta | \tilde{h}_{c})|^{2} \le 
  \sigma^{2} (\tilde{h}_{c} \delta \theta | \tilde{h}_{c} \delta \theta) \, .
\end{equation}
This guarantees that the magnitude of the second (positive) contribution
to the SNR from the phase error is always less than the first (negative)
one.  Hence, an error in the phase of the response function will always
serve to reduce the recovered SNR.

\subsubsection{Application to the LIGO instruments}
\label{sec:ligo_resp}

The calibration report from the LIGO S4 science run \cite{Dietz:2006}
provides an in depth analysis of the calibration of the LIGO
instruments.  Here, we briefly review those details which are relevant
for coalescing binary searches.

In the LIGO instruments, the response function is determined from the
sensing function $C(f)$ and the open loop gain function
$G(f)$ as
\begin{equation}
  R(f) =  \frac{ 1 + G(f)}{C(f)} \, ,
\end{equation}
These calibration functions, $C(f)$ and $G(f)$ are measured at intervals
during an analysis.  At intervening times, it is assumed that their
functional form is unchanged.  However, due to the changes in light
power stored in the arms, a time dependent rescaling $\gamma(t)$ of the
both the sensing function and open loop gain is required.  Thus, the
response function for any time can be expressed as:
\begin{equation}\label{eq:ligo_response}
  R(f,t) = \left( \frac{ 1 + \gamma(t) G_{o}(f) }{ \gamma(t) C_{o}(f) }
         \right) \, . 
\end{equation}

The uncertainties in the various components of the response function are
calculated in detail in Ref.~\cite{Dietz:2006}.  These can be summarized
by the statement that the uncertainties in the response function are of
order 5\% in the amplitude of the response function and $5^{\circ}$ in
phase.  Substituting this into our general expression
(\ref{eq:measured_snr}), we obtain:
\begin{equation}\label{eq:ligo_snr}
  \rho_{t} = \rho  
  \left[ 1 \mp \frac{|\delta r|}{r} - \frac{|\delta \theta |^{2}}{2} \right] 
  = \rho \left[ 1 \mp 0.05 \right] \, .
\end{equation}
Note that the error in the measured SNR is primarily due to the
amplitude calibration uncertainty.

\subsection{Uncertainties in the measured efficiency}
\label{ss:eff_uncertainties}

The effects of discreteness of the template placement,  errors in the
estimates of the power spectral density $S(f)$ used in the matched
filter in Eq.~(\ref{eq:snr}), and trends in the instrumental noise are
all accounted for by the Monte-Carlo simulation.  However, in the Monte
Carlo simulation, only a finite number of injections are performed as
these are computationally costly.  Thus, the efficiency plots, such as
that shown in Figure \ref{fig:sim_efficiency}, have an associated
statistical error.  The efficiency plots are produced by binning up the
parameter space and calculating the efficiency in each bin as
\begin{equation}
  \epsilon = \frac{N_{f}}{N_{f}+N_{m}} \, ,
\end{equation}
where $N_{f}$ and $N_{m}$ are the number of found and missed injections
respectively.  Then, assuming binomial errors for the efficiency, the
variance in the efficiency is:
\begin{equation}\label{eq:mc}
  \sigma^{2}_{\epsilon} = \frac{N_{f}N_{m}}{(N_{f} + N_{m})^{3}} 
  = \frac{ \epsilon ( 1 - \epsilon) }{(N_{f} + N_{m})}\, .
\end{equation}

As expected, the variance is inversely proportional to the number of
injections performed.  Furthermore, it is clear from (\ref{eq:mc}) that
the statistical uncertainty in the efficiency is greatest when the
efficiency is close to 50\%.  The Monte Carlo
error in the luminosity is obtained by multiplying the error in the
efficiency by the luminosity:%
\footnote{Strictly, there is a second Monte Carlo uncertainty due to
errors in the calculated luminosity.  However, the luminosity density of
Figure \ref{fig:luminosity} can be calculated to good accuracy with
minimal computational cost so that these errors will be insignificant.}
\begin{equation}\label{eq:lum_mc_error}
  \left(\Delta \cl(\rho)\right)^{2} = 
  \sum_{\deff, \mchirp} 
  \sigma^{2}_{\epsilon} (\rho, \mathbf{\deff}, \mchirp) \,
  L_{B}(\mathbf{\deff}) \, p(\mchirp)\, .
\end{equation}

When performing a search on real data, software injections are
computationally costly.  From Eq.~(\ref{eq:lum_mc_error}), we see that
simulations are most efficiently performed when concentrated where the
efficiency is close to $50\%$ and there is a significant contribution to
the luminosity.  Furthermore, for low-mass signals, making use of the
chirp distance (\ref{eq:dchirp}) reduces the dimension of
(\ref{eq:lum_mc_error}) by one and consequently reduces the size of the
associated Monte Carlo errors.

\subsection{Marginalization over Uncertainties to obtain an Upper Limit}
\label{sec:marginalization}

In section \ref{sec:upper_limit} we have described how to calculate an
upper limit from an inspiral search, making use of the loudest event
statistic.  This involves calculating three quantities, the amount of
time searched $T$, the total luminosity $\cl(\rhom)$ to which the search
is sensitive, and the likelihood ratio $\Lambda$ of the event being
foreground.  In this section, we have discussed various systematic
errors which affect our ability to measure these quantities.  Here, we
will describe how these uncertainties can be marginalized over to
produce an upper limit which takes them into account.

Let us begin by considering a single uncertainty (for example the
distance error) which will effect the luminosity $\cl$.  Then, by
evaluating errors associated to this uncertainty, we obtain a
probability distribution for the cumulative luminosity, $p_{d}(\cl)$.
In order to marginalize over the distance uncertainty, we simply
evaluate
\begin{equation}\label{eq:marginalize}
  p(R | \rhom,T,B) = \int d \cl \, p_{d}(\cl) \,  p(R | \rhom, T, B, \cl) 
  \, ,
\end{equation}
where the probability distributions on the right hand side must be
normalized to unity.  This is straightforward when there is only one
error to take into account.  However, in the preceding sections, we
have detailed several errors.  It is not practical to integrate over all
of these errors independently, so we perform a Monte Carlo integration.
For the majority of the errors, we use a Gaussian with standard
deviation given by the value stated above (and truncated so that the
total $L_{10}$ will never be negative).  Since the magnitude error
affects the luminosity exponentially, assuming a Gaussian error on this
leads to a log-normal distribution for the luminosity.  Finally, for the
waveform error, we use a 1-sided Gaussian, i.e. one which can only
decrease the cumulative luminosity.  Then, the Monte Carlo integral is
performed by sampling many times from the appropriate distributions.

\subsection{Uncertainties in the likelihood}
\label{sec:lambda}

We have cataloged various uncertainties which will affect the
calculated luminosity $\cl$ for a given search and shown how they can be
marginalized over to obtain a final distribution for the rate.  In
addition, we need to examine the effect any uncertainty in the
estimation of the likelihood $\Lambda$, will have upon the upper limit.
To do so, we will once again marginalize over this nuisance parameter
(as in Eq.~(\ref{eq:marginalize})), to obtain a distribution which is
independent of $\Lambda$.  However, due to the simple manner in which
$\Lambda$ enters the probability distribution (\ref{eq:rate_posterior})
for the rate, to leading order the marginalization procedure has no
effect on the distribution \cite{ul}.  More precisely, provided the
uncertainties in $\Lambda$ are small compared to $(1 + \Lambda)$, the
marginalization procedure serves to replace $\Lambda$ by its expectation
value.  We have argued that for a typical search where the loudest event
is consistent with the background, we expect $\Lambda \ll 1$ whence it
is safe to neglect uncertainties in its value.

There are cases when the estimated background from time shifts will not
accurately reflect the background.  An obvious example of this is when
computing the background for the co-located Hanford detectors.  It is
well known that there are correlated noise sources which will produce
inspiral triggers in the two instruments simultaneously.  While every
effort is made to remove these times by examination of auxiliary
channels, and in particular the seismic data, there are still invariably
some correlated noise event which occur in the Hanford data.  In this
case, time shifts will not give a good estimate of the background.
There are other ways to estimate the background, such as using ``reverse
chirp'' filters (obtained by time inverting the template).  However, in
calculating an upper limit, an underestimation of the background will
lead to a conservative upper limit being placed.

\subsection{An Example}
\label{sec:example}

To illustrate the issues associated with these systematic uncertainties,
we will calculate them for the example introduced earlier.  For our
example, the simulated loudest event had a combined SNR of $\rhom =
9.95$, which corresponds to a single instrument SNR around seven.  At
this SNR, the $50\%$ efficiency for BNS occurs at 40 Mpc.
The cumulative luminosity of such a search is $\cl(\rhom) = 540 \,
L_{10}$.  Finally, the value of the likelihood for our simulated results
is $\Lambda = 0.02$.  Now, we turn our attention to the systematic
uncertainties discussed above.  We will evaluate the effect of each of
these on the cumulative luminosity.

Distance uncertainties are obtained by moving all galaxies either closer
or further away, by the appropriate fraction given in the CBCG-catalog.
This changes the effective distance to sources according to
Eq.~(\ref{eq:delta_deff_dist}) and the luminosity according to
Eq.~(\ref{eq:luminosity_distance}).  This yields a change in cumulative
luminosity of $90 \, L_{10}$ with a luminosity decrease as galaxy distances
are increased.  Uncertainties in the magnitudes of galaxies are taken
into account by rescaling their luminosities according to
Eq.~(\ref{luminosity_magnitude}) and lead to an error in the cumulative
luminosity of $100 \, L_{10}$.  

Based on the discussion of Section \ref{sec:waveform_error}, for BNS
systems we choose to use a $10\%$ uncertainty in the astrophysical
waveform.  This is simulated by reducing the efficiency of both
detectors by the amount.  In other words, we rescale the axes in Figure
\ref{fig:2d_efficiency} downwards by $10\%$.  The effect in our
simulation is a change in the luminosity of $160 \, L_{10}$.  

The calibration uncertainty is calculated by varying the efficiency
curve accordingly, in a similar manner to that used for the waveform
errors.  Since the calibration errors in different instruments are
independent, we consider them one at a time.  In our example, we take a
$5\%$ calibration uncertainty in both the H1 and L1 detectors.  This is
calculated to lead to a $37 \,L_{10}$ variation in the luminosity in H1
and a $34 \,L_{10}$ variation in L1.  These numbers are very similar, as
expected.  They differ slightly due to the fact that certain galaxies
are better aligned for one detector than the other.

The Monte Carlo uncertainty can be calculated according to
Eq.~(\ref{eq:lum_mc_error}).  As an example, Figure
\ref{fig:sim_efficiency} was produced with 100 injections in each bin.
This gives a Monte Carlo error of $3 \, L_{10}$ which is already well
below the errors due to astrophysical and waveform uncertainties.

\begin{figure}
\includegraphics[width=1.0\linewidth]{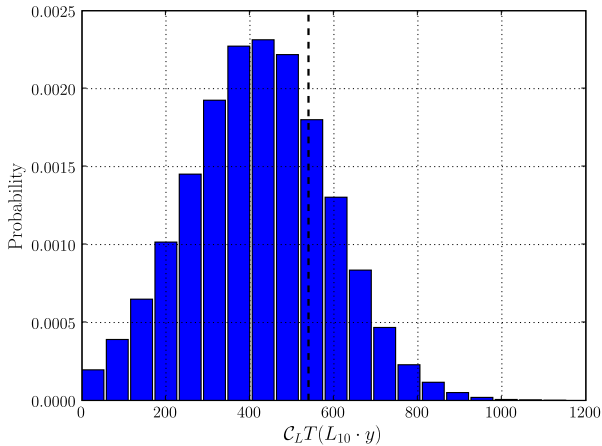}
\caption{\label{fig:luminosity_dist}
The distribution of the cumulative luminosity after marginalizing over
the systematic errors.  The histogram was produced by using 100,000
samples, and includes all the noise sources described in this section.
The dashed vertical line gives the luminosity before marginalization.
The fact that the peak of the distribution is below this is due to the
fact that the waveform errors are one sided.}
\end{figure}

The probability distribution for the cumulative luminosity taking into
account these systematic errors is shown in Figure
\ref{fig:luminosity_dist}.  To generate the distribution, the
uncertainties described above are used as the standard deviation of
Gaussian distributions sampled via a Monte Carlo process to obtain the
result.  As is clear from the figure, the width of the probability
distribution is significant.  The greatest contributions to the
uncertainty come from the astrophysics --- the uncertainty in the
luminosity distribution of nearby galaxies --- and uncertainties in the
waveform.  The systematic effects due to instrumental calibration and
Monte Carlo errors are sub-dominant.  

It is illustrative to consider the magnitude error and the waveform
error in more detail.  Although these errors are similar in magnitude,
the waveform error is one sided, and thus has a more significant effect
on the result.  Although the magnitude error is $100 \, L_{10}$,
marginalizing over it only increases the 90\% confidence upper limit
from $4.3 \times 10^{-3} \, L_{10}^{-1} yr^{-1}$ to $4.5 \times 10^{-3}
\, L_{10}^{-1} yr^{-1}$ .  In contrast, marginalizing over only the
waveform error increases the upper limit to $6.2 \times 10^{-3}
L_{10}^{-1} yr^{-1}$.  The effect is much more significant, even though
the magnitude of the two systematic errors is similar.  The reason is
that the waveform error can only decrease the sensitivity of the search.
Indeed, by modeling the waveform error as a 1-sided Gaussian, we are
significantly changing the mean value of $\cl(\rhom)$, reducing it from
$540 \, L_{10}$ to $410 \, L_{10}$, which accounts for a large fraction
of the increase in the reported upper limit.  When we include all of the
systematics together, we obtain a luminosity distribution as shown in
figure \ref{fig:luminosity_dist}.

\begin{figure}
\includegraphics[width=1.0\linewidth]{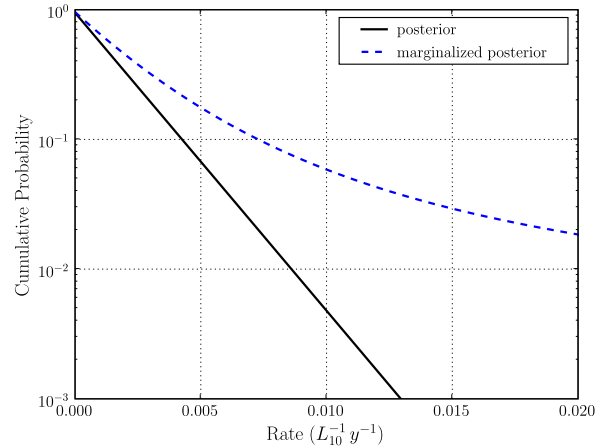}
\caption{\label{fig:posterior_cdf}
The cumulative distribution on the rate being greater than a given
value.  We are interested in the 90 \% upper limit and therefore take
the rate for which the distribution is equal to 0.1.  The distribution
is plotted for both the original luminosity of $540 L_{10}$ and the
marginalized luminosity distribution shown in figure
\ref{fig:luminosity_dist}.  The large tail on the marginalized rate is
due to the width of the luminosity distribution.}
\end{figure}

Finally, we can make use of the above distribution for the cumulative
luminosity in order to construct the posterior probability distribution
for the rate.  We do this both for the unmarginalized and marginalized
cases, and these are shown in Figure \ref{fig:posterior_cdf}.  This
shows the cumulative posterior probability distribution for the rate of
binary coalescence given the search.  The figure shows both the
un-marginalized and marginalized distributions.  The marginalized
distribution takes into account the systematic errors discussed above
and falls off more slowly due to the width of the luminosity
distribution from Figure \ref{fig:luminosity_dist}.  The final,
marginalized upper limit for this example can be read off from Figure
\ref{fig:posterior_cdf} as  $7.3 \times 10^{-3} \, L_{10}^{-1} \,
yr^{-1}$.

\section{Summary and Conclusions}
\label{sec:summary}

The astrophysical interpretation of the results is a critical part of
any search for gravitational waves.  In this paper, we have described a
method to obtain an astrophysical rate upper limit or interval from the
results of a search for coalescing binaries.  This method has been used
in obtaining the results from the LIGO S3 and S4 science runs
\cite{LIGOS3S4all}.  An astrophysical interpretation of a search result
requires a good understanding of both the detector sensitivity and the
relevant astrophysics.  We have argued that the detector sensitivity can
be expressed in terms of the efficiency and furthermore that, for
non-spinning systems, this efficiency is dependent only on the chirp
mass and effective distance of the binary relative to the detectors.
Furthermore, for low-mass systems, these can be combined into a single
quantity, the chirp distance (\ref{eq:dchirp}), which characterizes the
amplitude.  The use of effective distance is only appropriate for
non-spinning binaries.  Once spin is included, the orbital plane of the
binary precesses whence the effective distance is not a constant over
the course of inspiral.  Despite this, the methods presented here could
be extended to spinning binaries by making use of the ``expected SNR''
($\rho_{h}$ in Eq.~(\ref{eq:rho_h})) to characterize the amplitude of
the signal.  However, while it is clear that the efficiency for a
non-spinning binary will be a function only of the effective distance
and chirp mass, it is not obvious that the expected SNR and masses will
completely characterize the efficiency to spinning systems.  

The relevant astrophysics is encoded in the expected distribution of
coalescing binaries in the universe.  Following \cite{Phinney:1991ei},
we make the assumption that compact binaries are distributed according
to the blue light luminosity.  Making use of a catalog of nearby
galaxies (such as the one in Ref.~\cite{LIGOS3S4Galaxies}), we obtained
an expression for the total luminosity to which a given search is
sensitive.  The loudest event statistic allows us to obtain a
probability distribution for the rate of binary coalescence given the
results of a search.  This distribution depends upon the cumulative
luminosity discussed above as well as an understanding of the rate of
background, noise events present in the data.  Finally, the posterior
distribution for the rate can be used to calculate an upper limit or
rate interval for the occurrence of binary coalescence.

There are numerous systematic uncertainties involved in the calculation
of the rate which we have discussed in detail.  The dominant errors
arise due to uncertainties in the distribution of nearby galaxies and
imprecise knowledge of the gravitational waveform emitted by coalescing
binaries.  It is reassuring that the errors associated with our
understanding of the detectors and the analysis, such as calibration and
Monte Carlo statistical uncertainties, are less significant than the
physical and astrophysical uncertainties discussed above.  

Uncertainties in the measurement of distances and apparent magnitudes of
nearby galaxies leads to an uncertainty in the total luminosity
available to a search.  This in turn affects the reported rate limit.
Additionally, the unknown mass distribution of coalescing binaries will
have a significant effect on the reported rate.  To mitigate this
effect, rates are reported as a function of mass.  In the future,
gravitational-wave observations of binaries by advanced detectors will
provide improved knowledge of both the mass and location distribution of
binaries.  Although the post--Newtonian waveform is known to a high
level of accuracy, uncertainties in the waveform, the neglect of spin
and amplitude corrections lead to significant uncertainties in the
sensitivity of the search and hence the reported rate.  In future
searches, it should be possible to quantify the waveform uncertainties
more precisely by performing simulations with amplitude corrected,
spinning waveforms and by comparing the post--Newtonian waveforms to
those obtained from numerical relativity simulations.

\section*{Acknowledgements}

We would like to acknowledge many useful discussions with members of
the LIGO Scientific Collaboration inspiral analysis group which were
critical in the formulation of the methods and results described in
this paper. This work has been supported in part by NSF grant
PHY-0200852, a Cottrell Scholar Award from the Research Corporation
(PRB), and the Royal Society (SF). LIGO was constructed by the
California Institute of Technology and Massachusetts Institute of
Technology with funding from the National Science Foundation and
operates under cooperative agreement PHY-0107417.  This paper has LIGO
Document Number LIGO-P070077-00-Z.

\bibliography{../bibtex/iulpapers}

\end{document}